\documentclass{PoS}

\usepackage{microtype}
\usepackage{amsmath}
\usepackage{sidecap}

\title{Nucleon structure with pion mass down to 149 MeV}

\ShortTitle{Nucleon structure with pion mass down to 149 MeV}

\author{\speaker{Jeremy~Green},$^a$ Michael~Engelhardt,$^b$
  Stefan~Krieg,$^{cd}$ John~Negele,$^a$ Andrew~Pochinsky$^a$
  and Sergey~Syritsyn$^e$\\
  \llap{$^a$}Center for Theoretical Physics, Massachusetts Institute of Technology,\\
  Cambridge, Massachusetts 02139, USA\\
  \llap{$^b$}Department of Physics, New Mexico State University,
  Las Cruces, New Mexico 88003, USA\\
  \llap{$^c$}Bergische Universit\"at Wuppertal,
  D-42119 Wuppertal, Germany\\
  \llap{$^d$}IAS, J\"ulich Supercomputing Centre, Forschungszentrum J\"ulich,
  D-52425 J\"ulich, Germany\\
  \llap{$^e$}Lawrence Berkeley National Laboratory,
  Berkeley, California 94720, USA\\
  E-mail: \email{jrgreen@mit.edu}, \email{engel@physics.nmsu.edu},
  \email{s.krieg@fz-juelich.de}, \email{negele@mit.edu}, \email{avp@mit.edu},
  \email{ssyritsyn@lbl.gov}}

\abstract{We present isovector nucleon observables: the axial, tensor,
  and scalar charges and the Dirac radius. Using the BMW
  clover-improved Wilson action and pion masses as low as 149~MeV, we
  achieve good control over chiral extrapolation to the physical
  point. Our analysis is done using three different source-sink
  separations in order to identify excited-state effects, and we make
  use of the summation method to reduce their size.}

\FullConference{The 30th International Symposium on Lattice Field Theory\\
		 June 24 -- 29,  2012\\
		 Cairns, Australia}

\begin{document}

\section{Introduction}

Lattice QCD calculations of nucleon structure observables are now
entering the stage where they are being seriously confronted with
experiment, although resources are not sufficient for full control
over all systematic errors. Thus, when predicting unobserved
quantities, it is useful to know which sources of systematic error
must be more carefully controlled in order to achieve agreement for
experimentally observed quantities, and which sources of error are
already well under control when using standard techniques.

We report on two experimentally measured observables, the isovector
Dirac radius $(r_1^2)^v$ and the axial charge $g_A$; and two
predictions, the tensor and scalar charges, $g_T$ and $g_S$. Proton
matrix elements of the vector current are parameterized by two form
factors,
\begin{equation}
  \langle p(P')|\bar q\gamma^\mu q|p(P)\rangle = \bar u(P')\left( \gamma^\mu F_1^q(Q^2) + \tfrac{i\sigma^{\mu\nu}\Delta_\nu}{2m} F_2^q(Q^2) \right) u(P),
\end{equation}
where $\Delta=P'-P$ and $Q^2=-\Delta^2$. The isovector Dirac radius is
defined from the slope of the isovector $F_1^v\equiv F_1^u-F_1^d$ at
zero momentum transfer:
\begin{equation}
  F_1^v(Q^2) = F_1^v(0)(1-\tfrac{1}{6}(r_1^2)^vQ^2 + \mathcal{O}(Q^4)).
\end{equation}
In terms of experimental observables, $(r_1^2)^v$ is related to the
difference between proton and neutron charge radii, the former of
which has a $7\sigma$ discrepancy between measurements from
electron-proton scattering~\cite{Beringer:1900zz}
and a recent result using a precise measurement of the Lamb shift in
muonic hydrogen~\cite{Pohl:2010zza}. Lattice QCD calculations with a
precision of a few percent could help to resolve this.

The axial, tensor, and scalar charges all have similar definitions
from neutron-to-proton transition matrix elements at zero momentum
transfer:
\begin{equation}
  \langle p|\bar u\gamma^\mu\gamma_5 d|n\rangle = g_A \bar u_p \gamma^\mu\gamma_5 u_n, \qquad
  \langle p|\bar u\sigma^{\mu\nu} d|n\rangle = g_T \bar u_p \sigma^{\mu\nu} u_n, \qquad
  \langle p|\bar u d|n\rangle = g_S \bar u_p u_n.
\end{equation}

The axial charge is a key benchmark observable, since it is a
naturally isovector quantity (and thus not requiring disconnected
diagrams to calculate it), measured via forward matrix elements, and
it is also well-measured experimentally via beta decay of polarized
neutrons.

The tensor and scalar charges have not been measured experimentally,
but it has recently been shown~\cite{Bhattacharya:2011qm} that they
control the leading contributions to neutron beta decay from new
(beyond the Standard Model) physics, and thus they provide a useful
input to the analysis of experimental data.

\section{Methodology}

\begin{SCfigure}
  \centering
  \includegraphics[width=0.42\columnwidth]{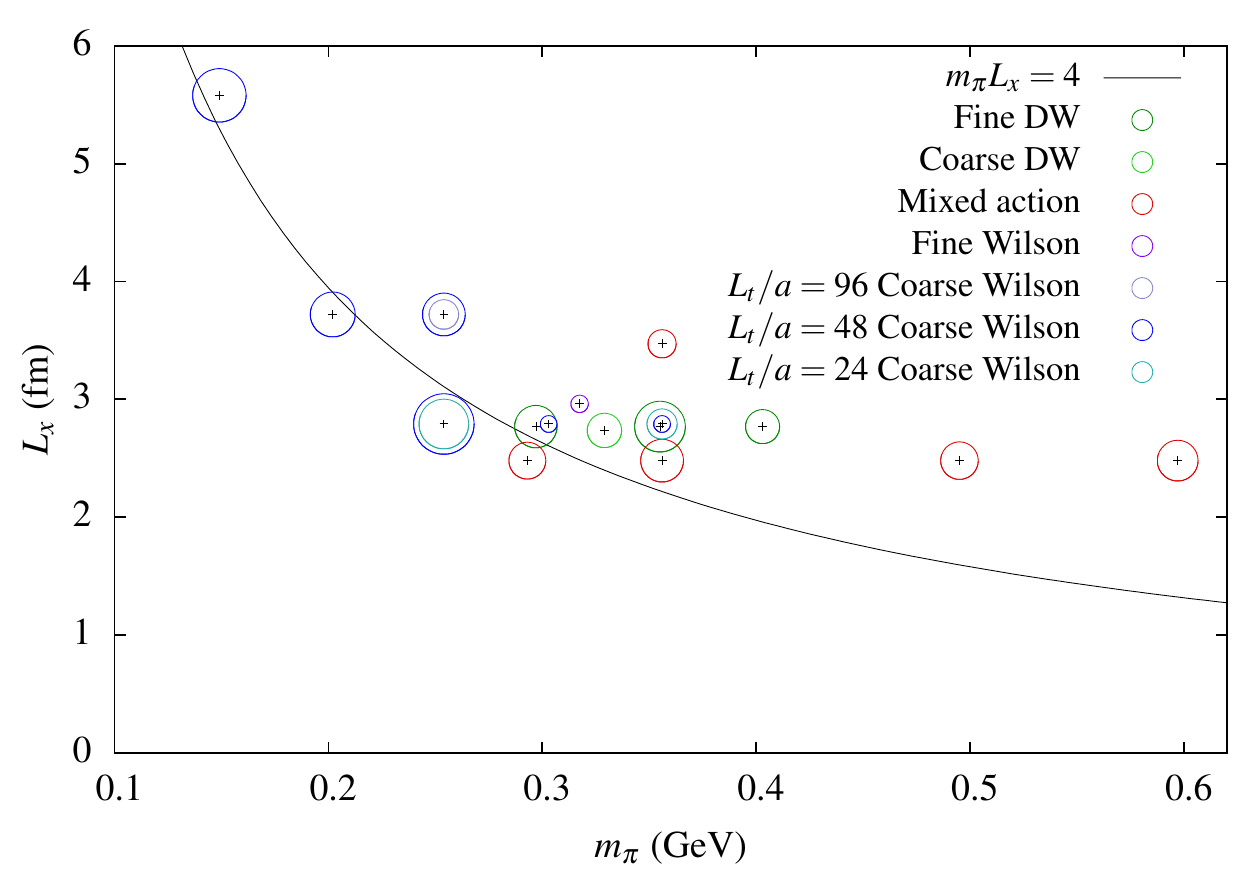}
  \caption{Spatial box length $L_x$ and pion mass for the full set of
    ensembles. The areas of the circles are proportional to the number
    of measurements made on each ensemble.}
  \label{fig:ensembles}
\end{SCfigure}

The main results presented are from calculations performed on ten
Lattice QCD ensembles using $2+1$~flavors of tree-level
clover-improved Wilson fermions coupled to double HEX-smeared gauge
fields~\cite{Durr:2010aw}. We also compare with results from earlier
$2+1$~flavor
calculations~\cite{Syritsyn:2009mx,Bratt:2010jn,Syritsyn_thesis} using
four ensembles with unitary domain wall
fermions~\cite{Allton:2008pn,Aoki:2010dy}, as well as five ensembles
using a mixed-action scheme with domain wall valence quarks and Asqtad
staggered sea quarks~\cite{Bernard:2001av}. The ranges of parameters
used in these calculations are summarized in Tab.~\ref{tab:actions}
and Fig.~\ref{fig:ensembles}.

\begin{table}
  \centering
  \begin{tabular}{llccccc}
    Action & $a$ (fm) & $m_\pi$ (MeV) & $L_x/a$ & $L_t/a$ & $T/a$ & \# meas \\\hline
    Wilson & 0.09 & 317(2) & 32 & 64 & 10, 13, 16 & 824\\
    Wilson & 0.116 & 149--356 & 24, 32, 48 & 24, 48, 96 & ~8, 10, 12 & 762--10032 \\
    Domain wall & 0.084 & 297--403 & 32 & 64 & 12 & 4216--7056 \\
    Domain wall & 0.114 & 329(5) & 24 & 64 & 9 & 3192 \\
    Mixed & 0.124 & 293--597 & 20, 28 & 64 & 9 & 2176--5024
  \end{tabular}
  \caption{Lattice actions and ranges of parameters used: lattice spacing,
    pion mass, spatial and temporal box size, 
    source-sink separation, and number of measurements.}
  \label{tab:actions}
\end{table}

On every ensemble with Wilson fermions, we compute nucleon three-point
functions using three different source-sink separations $T$. When
matrix elements are computed using the traditional ratio-plateau
method, the asymptotically dominant excited-state contaminations arise
from transitions between the ground state and the lowest excited
state, and these decay as $e^{-\Delta E T/2}$. The summation method
requires combining more than one source-sink separation, but yields
improved asymptotic behavior, with the leading contaminants to forward
matrix elements decaying as $Te^{-\Delta E
  T}$~\cite{Capitani:2010sg,Bulava:2010ej}.

This set of ensembles allows for control over, or study of, various
sources of systematic error. In roughly decreasing order of the level
of control that we can achieve:
\begin{description}
\item[Quark masses] For all ensembles, the strange quark mass is near
  the physical value. Our smallest $m_{ud}$ corresponds to a pion mass
  of 149~MeV, which is just 10\% above the physical pion mass. This
  allows for either a direct comparison of the $m_\pi=149$~MeV
  ensemble with experiment, or a comparison after a mild chiral
  extrapolation to the physical pion mass.
\item[Excited states] Using the ratio-plateau method with three
  different source-sink separations allows for clear identification of
  observables where excited-state contamination is a problem. These
  three source-sink separations can also be combined using the
  summation method to get another result that may be less affected by
  excited states.
\item[Finite volume] In general, finite-volume effects are expected to
  be small with $m_\pi L\gtrsim 4$. Furthermore, we can perform controlled
  comparisons between the $24^3\times 48$ and $32^3\times 48$ Wilson
  ensembles near $m_\pi=250$~MeV, where the spatial volume is changed
  while other parameters are fixed.
\item[Finite temperature] At our smallest pion masses, the lattice
  time extent is shorter than the typically used $L_t=2L_x$, and these
  ensembles may be particularly susceptible to thermal effects. On the
  other hand, the three different time extents $L_t/a\in\{24,48,96\}$
  used for Wilson ensembles near $m_\pi=250$~MeV are useful for
  identifying possible problems.
\item[Discretization] The use of different lattice actions and
  different lattice spacings allows for a consistency check, but this
  set of ensembles is insufficient for taking a continuum limit.
\end{description}

\section{Results}

For each of the observables, two plots are shown. The first includes
the data from all three actions. In order to show results using
similar techniques, the middle source-sink separation on the Wilson
action enembles is shown for this comparison. The second plot shows
the dependence on source-sink separation and the summation result for
each of the ten Wilson ensembles.

\begin{figure}
  \centering
  \includegraphics[width=0.495\columnwidth]{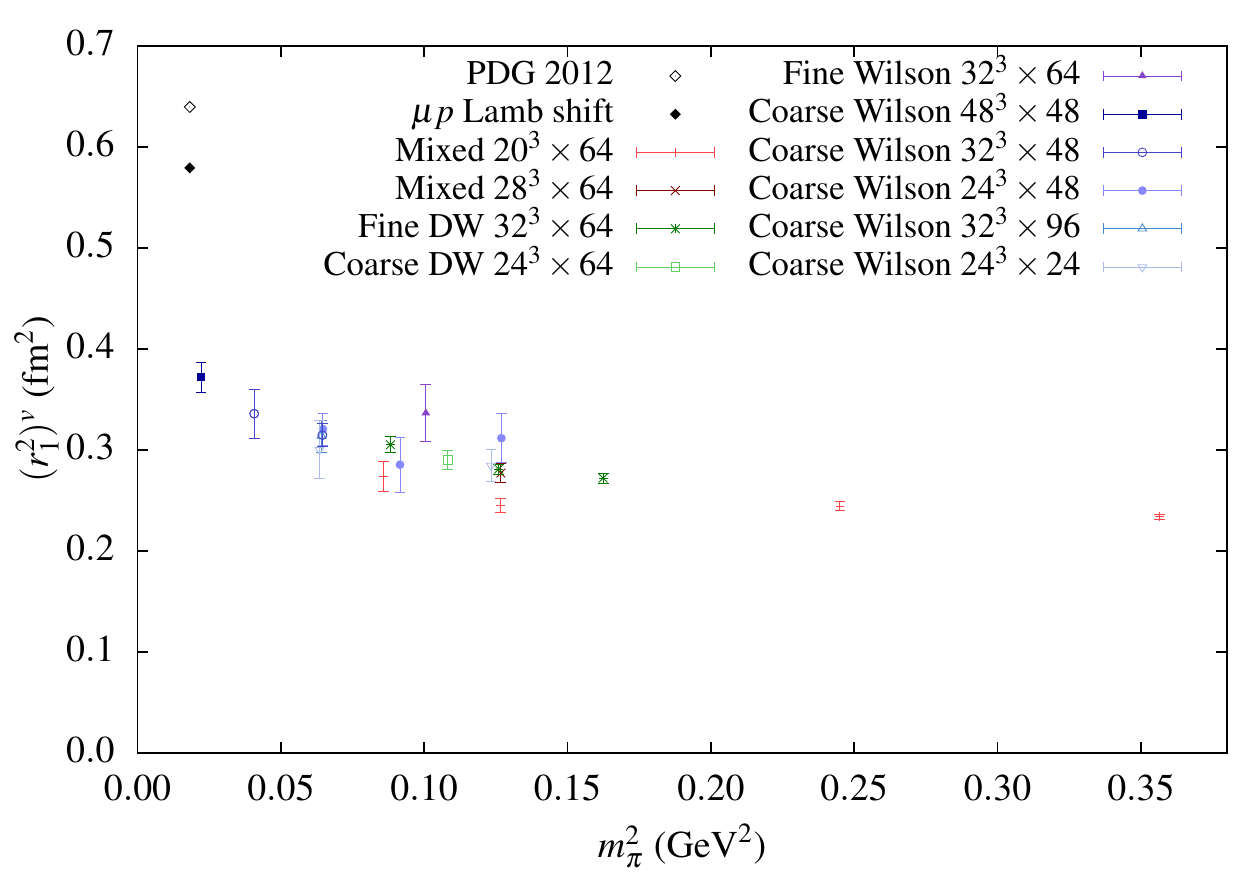}
  \includegraphics[width=0.495\columnwidth]{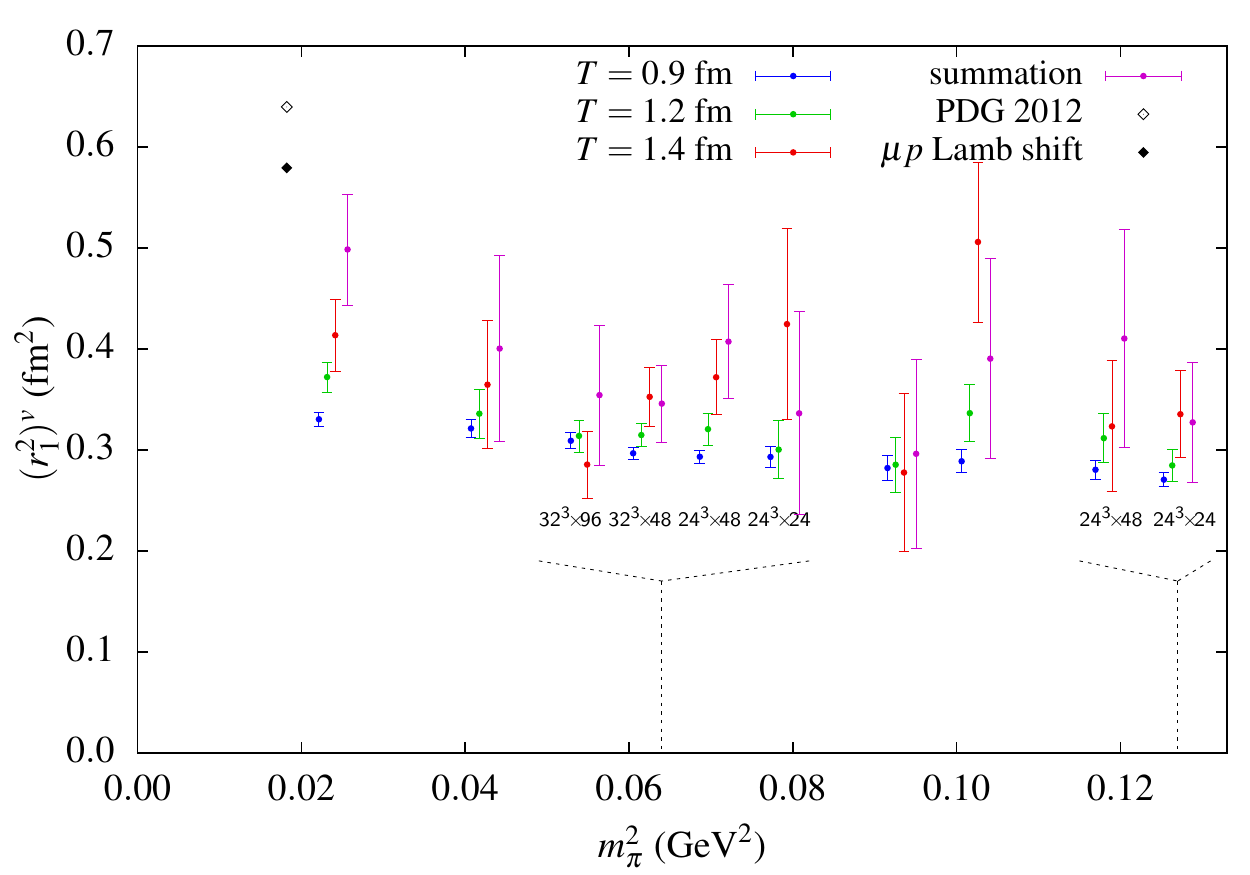}
  \caption{Isovector Dirac radius $(r_1^2)^v$, as determined from
    dipole fits to $F_1(Q^2)$. The two experimental points both use
    the PDG~\cite{Beringer:1900zz} value for $(r_E^2)^n$, and
    $(r_E^2)^p$ is taken from either the PDG or from the result from
    measurement of the Lamb shift in muonic
    hydrogen~\cite{Pohl:2010zza}. \textbf{Left:} Results from
    ensembles using the three different lattice actions. Wilson action
    points are taken from the middle source-sink
    separation. \textbf{Right:} The full set of Wilson action
    results. Measurements using the three source-sink separations and
    the summation method are slightly displaced horizontally. The
    points corresponding to the smallest source-sink separation are
    placed at the measured value of $m_\pi^2$, except for the ensembles
    with $m_\pi\approx 250$~MeV and $m_\pi\approx 350$~MeV, where the
    different volumes are displaced horizontally and the dotted lines
    indicate the approximate measured values of $m_\pi^2$. }
  \label{fig:r1v2}
\end{figure}

Figure~\ref{fig:r1v2} shows the isovector Dirac radius. This is
extracted from measurements of the isovector Dirac form factor
$F_1^v(Q^2)$ via dipole fits $F_1^v(Q^2)\sim
\frac{F_1^v(0)}{(1+Q^2/M_D^2)^2}$. The first plot shows a broad
consistency among the three lattice actions in their region of
overlapping pion masses, although the unitary domain wall data are
systematically a bit higher than the mixed action data. There is a
gentle rise as $m_\pi$ approaches the physical value, but all of these
data significantly undershoot the experimental results. The second
plot shows that excited-state effects are responsible for a large part
of this discrepancy with experiment. There is a general trend for the
data on each ensemble to increase with source-sink separation, and on
the lightest ensemble the summation point is near the experimental
points.  Chiral extrapolation of the summation data to the physical
pion mass yields a result consistent with
experiment~\cite{Green:2012ud}. As shown in Ref.~\cite{Green:2012ud},
calculations of the isovector Pauli radius, magnetic moment, and quark
momentum fraction behave similarly: results for the 149~MeV ensemble
monotonically approach the experimental value with increasing
source-sink separation, the summation point agrees with experiment,
and a chiral extrapolation of the summation data to the physical pion
mass yields results in agreement with experiment and having a smaller
statistical uncertainty than the 149~MeV summation point.  Further
study will be needed to obtain full control over excited-state effects, but
with presently available data we consider the summation method
as our best approach for reducing their size, and it succeeds in producing
agreement with experiment for these four observables. Comparing the
summation data on the four different space-time volumes at
$m_\pi\approx 250$~MeV, we see that there is no statistically
significant dependence on spatial volume or time extent. The summation
point on the fine ensemble at $m_\pi=317$~MeV is also consistent with
those on the nearby coarse ensembles, indicating the absence of
discretization effects at this level of precision.

\begin{figure}
  \centering
  \includegraphics[width=0.495\columnwidth]{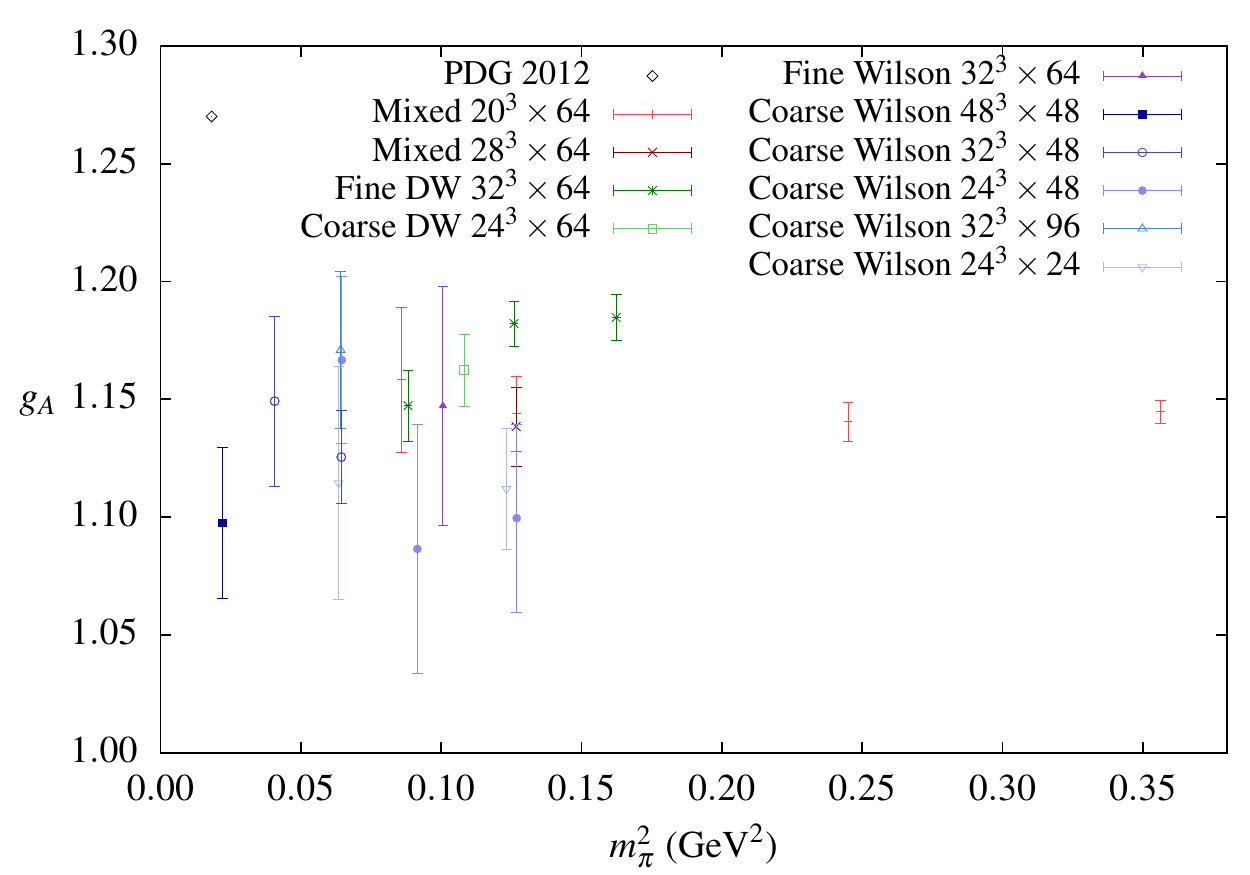}
  \includegraphics[width=0.495\columnwidth]{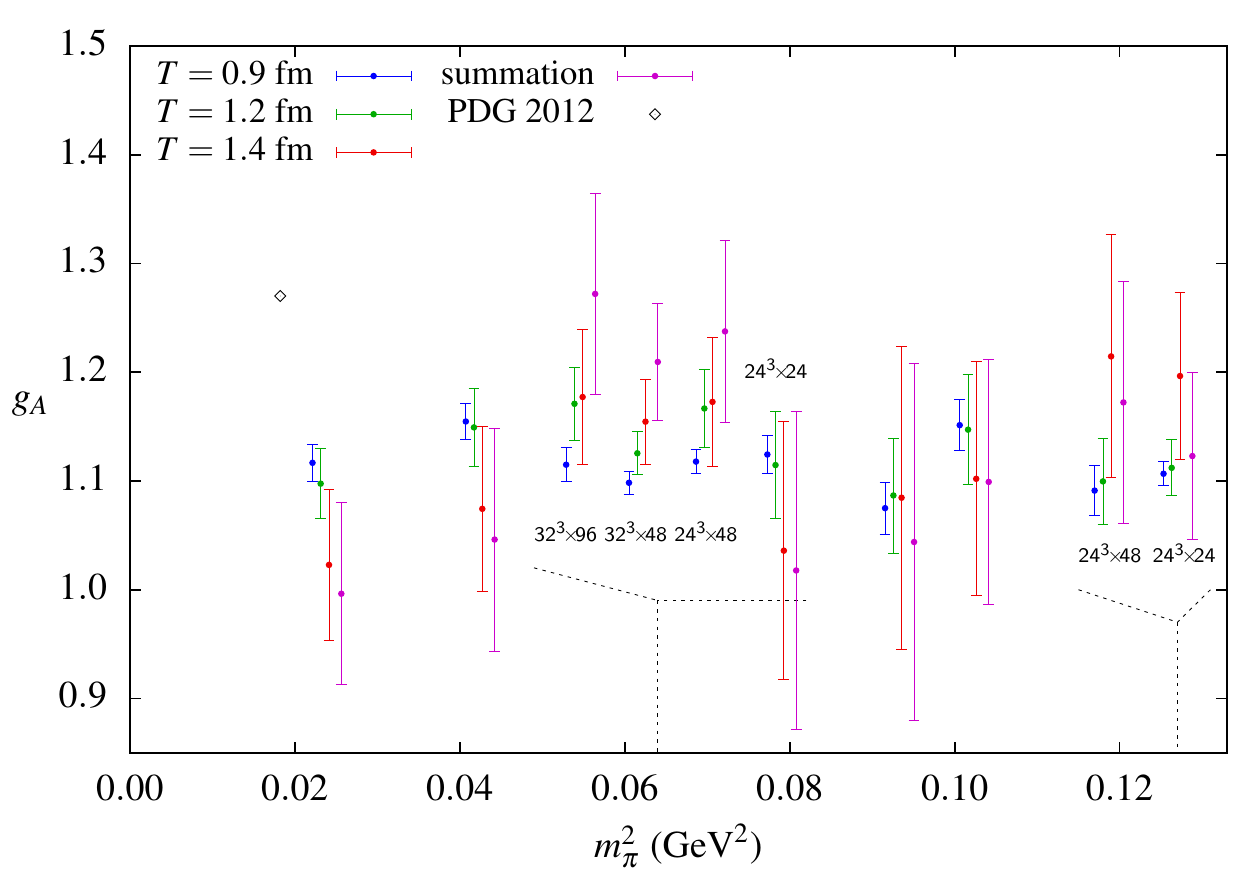}
  \caption{Axial charge $g_A$. The experimental value is from
    Ref.~\cite{Beringer:1900zz}. See caption of
    Fig.~\protect\ref{fig:r1v2}.}
  \label{fig:gA}
\end{figure}

The axial charge $g_A$ is shown in Fig.~\ref{fig:gA}. Again there is
broad agreement among the different actions. There is no clear
dependence on $m_\pi$, and (using the middle source-sink separation
for the Wilson ensembles) the data undershoot experiment by about
10\%. Looking at the second plot, we see that, on the lightest two
ensembles, increasing the source-sink separation moves the data away
from experiment. The opposite behavior is seen in a subset of the four
$m_\pi\approx 250$~MeV ensembles. For the three with $L_t/a\geq 48$,
increasing the source-sink separation moves the data toward experiment
and the summation points are consistent with experiment. In contrast,
the fourth with $L_t/a=24$ behaves similarly to the lightest two
ensembles, albeit with larger statistical uncertainty. This suggests
that the decrease of $g_A$ with source-sink separation is caused by
the influence of thermal pion states, since the three ensembles that
show this behavior all have small $m_\pi L_t$.

\begin{figure}
  \centering
  \includegraphics[width=0.495\columnwidth]{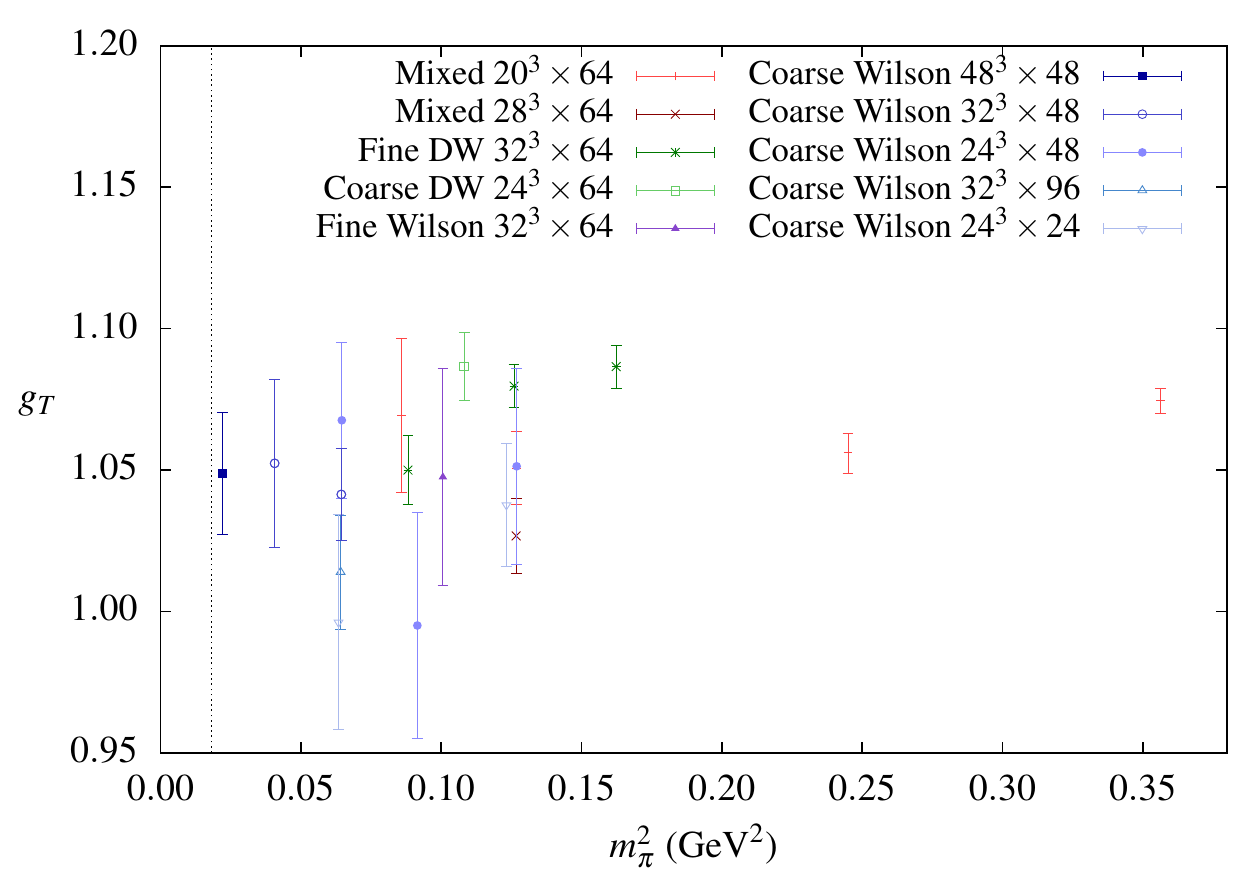}
  \includegraphics[width=0.495\columnwidth]{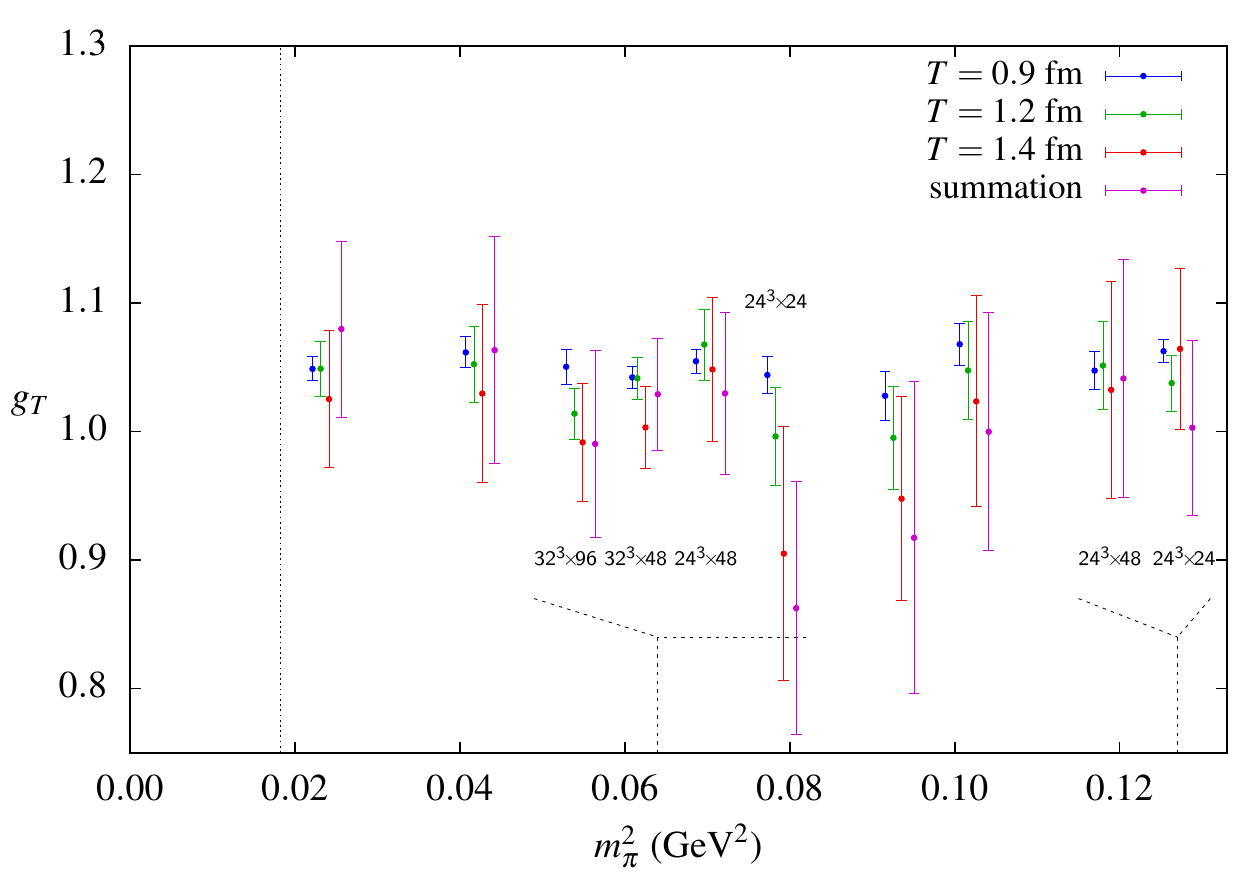}
  \caption{Tensor charge $g_T$. The physical pion mass is indicated by
    the vertical line. See caption of Fig.~\protect\ref{fig:r1v2}.}
  \label{fig:gT}
\end{figure}

\begin{figure}
  \centering
  \includegraphics[width=0.495\columnwidth]{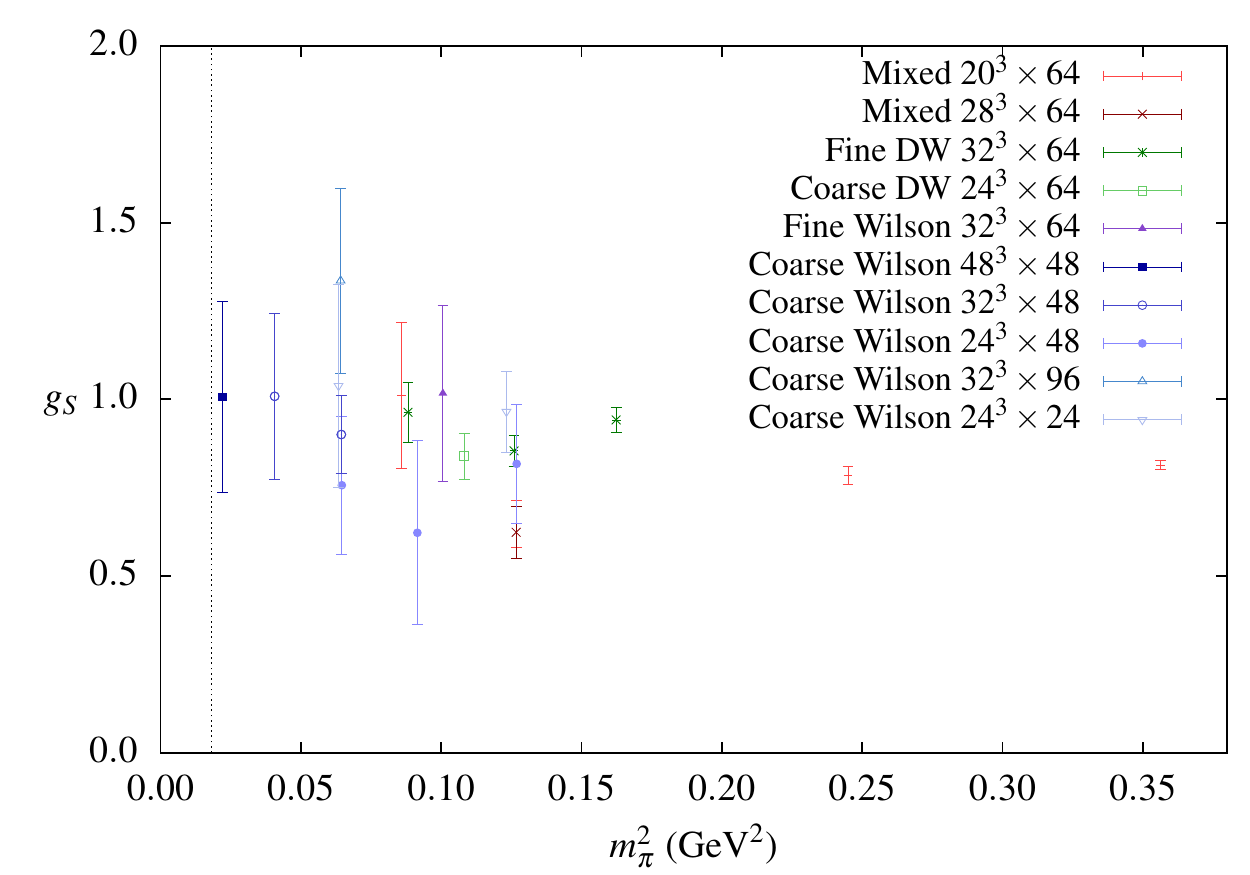}
  \includegraphics[width=0.495\columnwidth]{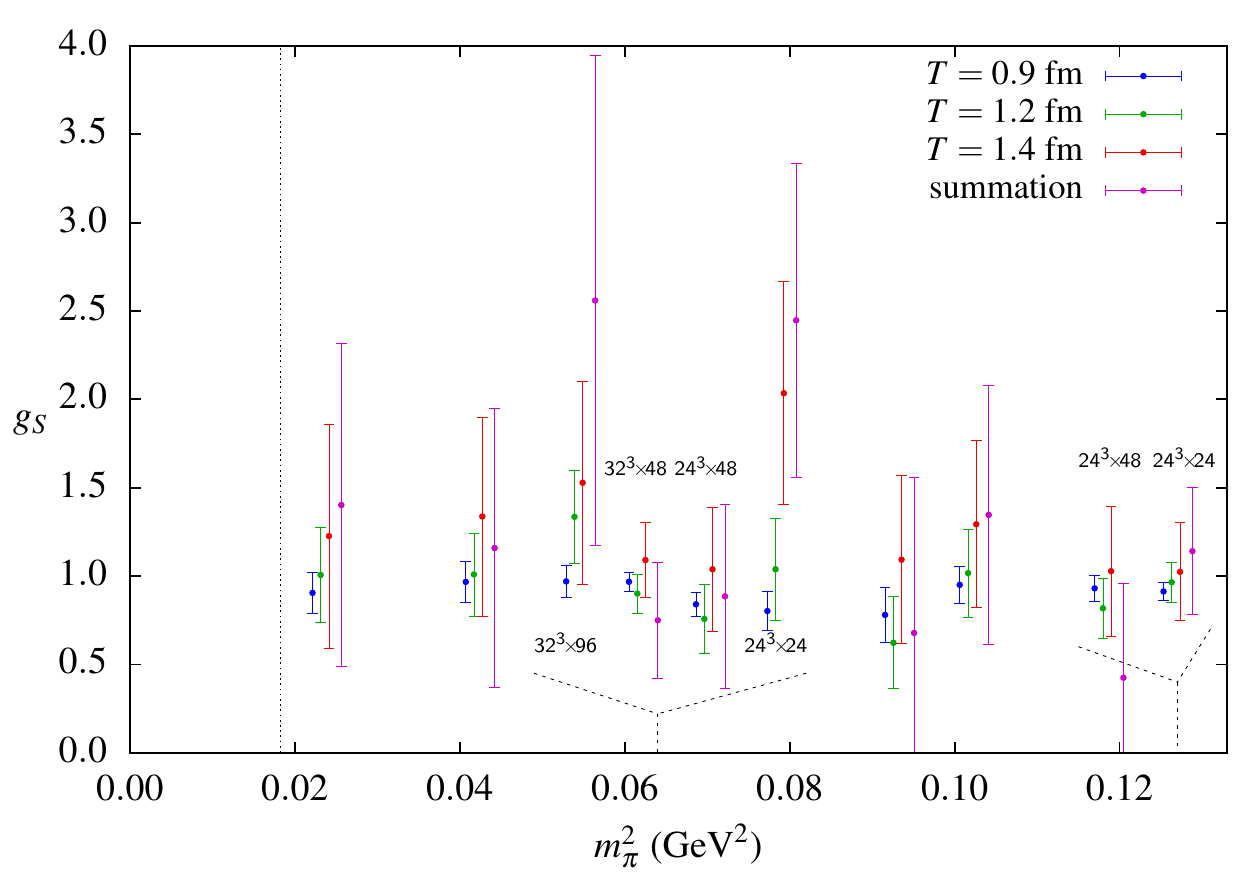}
  \caption{Scalar charge $g_S$. The physical pion mass is indicated by
    the vertical line. See caption of Fig.~\protect\ref{fig:r1v2}.}
  \label{fig:gS}
\end{figure}

Figures~\ref{fig:gT} and~\ref{fig:gS} show the tensor charge $g_T$ and
the scalar charge $g_S$, respectively. Measurements of the latter are
much noisier; note the significantly larger range of the vertical axis
in Fig.~\ref{fig:gS} compared with Figs.~\ref{fig:gA}
and~\ref{fig:gT}. Neither $g_S$ nor $g_T$ shows a clear statistically
significant dependence on source-sink separation, so using the middle
source-sink separation should give better results than was the case
for $(r_1^2)^v$ and $g_A$. At $m_\pi\approx 250$~MeV, there isn't a
significant dependence on the spatial volume, but the ensemble with
small $L_t$ shows a dependence on source-sink separation for both
$g_S$ and $g_T$. This behavior isn't seen on the ensembles with the
smallest pion masses, so it is likely that they aren't strongly
affected by thermal states. From the middle source-sink separation on
the $m_\pi=149$~MeV ensemble, we get $g_S=1.01(27)$ and
$g_T=1.049(23)$. Extrapolation to the physical pion mass using either
the coarse Wilson ensembles, or a global fit to all ensembles, yields
results consistent with the lightest ensemble~\cite{Green:2012ej}.

\section{Conclusions}

The importance of near-physical quark masses for nucelon structure
calculations is illustrated by the isovector Dirac radius, where the
rise toward experiment is only seen at our lightest pion masses, and
for the axial charge, where new behavior was seen only below
$m_\pi\approx 250$~MeV. In addition, it is essential that
excited-state effects can be identified, as shown clearly for the
isovector Dirac radius. This shows up even more dramatically for the
isovector quark momentum fraction $\langle
x\rangle_{u-d}$~\cite{Green:2011fg,Green:2012ud}. More study of
excited-state effects is required, and this may require large
computing resources, since as the source-sink separation $T$ is
increased to reduce excited-state contamination, the signal-to-noise
ratio is expected to decay as
$e^{-(m_N-\frac{3}{2}m_\pi)T}$~\cite{Lepage:1989hd}. We have
identified finite-temperature effects as a possible source of the
discrepancy with experiment for the axial charge.

As we obtain a better understanding of systematic errors, predictions
of nucleon properties using Lattice QCD become more
credible. Calculations of the nucleon scalar and tensor charge will
provide useful input to searches for new physics.

\acknowledgments

We thank Zoltan Fodor for useful discussions and the
Budapest-Marseille-Wuppertal collaboration for making some of their
configurations available to us.  This research used resources of the
Argonne Leadership Computing Facility at Argonne National Laboratory,
which is supported by the Office of Science of the U.S. Department of
Energy under contract \#DE--AC02--06CH11357, resources provided by the
New Mexico Computing Applications Center (NMCAC) on Encanto, resources
at Forschungszentrum J\"ulich, and facilities of the USQCD
Collaboration, which are funded by the Office of Science of the
U.S.~Department of Energy.

During this research JG, SK, JN, AP, and SS were supported in part by the
U.S.~Department of Energy Office of Nuclear Physics under grant
\#DE--FG02--94ER40818, ME was supported in part by DOE grant
\#DE--FG02--96ER40965, SS was supported in part by DOE contract
\#DE--AC02--05CH11231, and SK was supported in part by Deutsche
Forschungsgemeinschaft through grant SFB--TRR~55.

The Chroma software suite \cite{Edwards:2004sx} was used for the mixed
action and unitary domain wall calculations. The Wilson-clover
calculations were performed with Qlua~\cite{Qlua}.

\bibliographystyle{JHEP-2-notitle}
\bibliography{nucleonstructure.bib}

\end{document}